\PassOptionsToPackage{unicode}{hyperref}
\PassOptionsToPackage{hyphens}{url}
\PassOptionsToPackage{dvipsnames,svgnames,x11names}{xcolor}
\documentclass[
  12pt,
  letterpaper,
]{article}

\usepackage{amsmath,amssymb}
\usepackage{setspace}
\usepackage{iftex}
\ifPDFTeX
  \usepackage[T1]{fontenc}
  \usepackage[utf8]{inputenc}
  \usepackage{textcomp} 
\else 
  \usepackage{unicode-math}
  \defaultfontfeatures{Scale=MatchLowercase}
  \defaultfontfeatures[\rmfamily]{Ligatures=TeX,Scale=1}
\fi
\usepackage{lmodern}
\ifPDFTeX\else  
\fi
\IfFileExists{upquote.sty}{\usepackage{upquote}}{}
\IfFileExists{microtype.sty}{
  \usepackage[]{microtype}
  \UseMicrotypeSet[protrusion]{basicmath} 
}{}
\makeatletter
\@ifundefined{KOMAClassName}{
  \IfFileExists{parskip.sty}{%
    \usepackage{parskip}
  }{
    \setlength{\parindent}{0pt}
    \setlength{\parskip}{6pt plus 2pt minus 1pt}}
}{
  \KOMAoptions{parskip=half}}
\makeatother
\usepackage{xcolor}
\usepackage[top=0.75in,bottom=0.75in,left=1in,right=1in]{geometry}
\makeatletter
\ifx\paragraph\undefined\else
  \let\oldparagraph\paragraph
  \renewcommand{\paragraph}{
    \@ifstar
      \xxxParagraphStar
      \xxxParagraphNoStar
  }
  \newcommand{\xxxParagraphStar}[1]{\oldparagraph*{#1}\mbox{}}
  \newcommand{\xxxParagraphNoStar}[1]{\oldparagraph{#1}\mbox{}}
\fi
\ifx\subparagraph\undefined\else
  \let\oldsubparagraph\subparagraph
  \renewcommand{\subparagraph}{
    \@ifstar
      \xxxSubParagraphStar
      \xxxSubParagraphNoStar
  }
  \newcommand{\xxxSubParagraphStar}[1]{\oldsubparagraph*{#1}\mbox{}}
  \newcommand{\xxxSubParagraphNoStar}[1]{\oldsubparagraph{#1}\mbox{}}
\fi
\makeatother

\providecommand{\tightlist}{%
  \setlength{\itemsep}{0pt}\setlength{\parskip}{0pt}}\usepackage{longtable,booktabs,array}
\usepackage{calc} 
\usepackage{etoolbox}
\makeatletter
\patchcmd\longtable{\par}{\if@noskipsec\mbox{}\fi\par}{}{}
\makeatother
\IfFileExists{footnotehyper.sty}{\usepackage{footnotehyper}}{\usepackage{footnote}}
\makesavenoteenv{longtable}
\usepackage{graphicx}
\makeatletter
\newsavebox\pandoc@box
\newcommand*\pandocbounded[1]{
  \sbox\pandoc@box{#1}%
  \Gscale@div\@tempa{\textheight}{\dimexpr\ht\pandoc@box+\dp\pandoc@box\relax}%
  \Gscale@div\@tempb{\linewidth}{\wd\pandoc@box}%
  \ifdim\@tempb\p@<\@tempa\p@\let\@tempa\@tempb\fi
  \ifdim\@tempa\p@<\p@\scalebox{\@tempa}{\usebox\pandoc@box}%
  \else\usebox{\pandoc@box}%
  \fi%
}
\def\fps@figure{htbp}
\makeatother
\NewDocumentCommand\citeproctext{}{}
\NewDocumentCommand\citeproc{mm}{%
  \begingroup\def\citeproctext{#2}\cite{#1}\endgroup}
\makeatletter
 \let\@cite@ofmt\@firstofone
 \def\@biblabel#1{}
 \def\@cite#1#2{{#1\if@tempswa , #2\fi}}
\makeatother
\newlength{\cslhangindent}
\newlength{\csllabelwidth}
\newenvironment{CSLReferences}[2] 
 {\begin{list}{}{%
  \setlength{\itemindent}{0pt}
  \setlength{\leftmargin}{0pt}
  \setlength{\parsep}{0pt}
  \ifodd #1
   \setlength{\leftmargin}{\cslhangindent}
   \setlength{\itemindent}{-1\cslhangindent}
  \fi
  \setlength{\itemsep}{#2\baselineskip}}}
 {\end{list}}
\usepackage{calc}

\newcommand{\CSLLeftMargin}[1]{\parbox[t]{\csllabelwidth}{\strut#1\strut}}
\newcommand{\CSLRightInline}[1]{\parbox[t]{\linewidth - \csllabelwidth}{\strut#1\strut}}

\usepackage{algorithm}
\usepackage{algpseudocode}
\usepackage{graphicx}
\usepackage{scrlayer-scrpage}
\usepackage{etoolbox}
\usepackage{setspace}
\clearpairofpagestyles
\ofoot*{\pagemark}
\makeatletter
\@ifpackageloaded{caption}{}{\usepackage{caption}}
\AtBeginDocument{%
\ifdefined\contentsname
  \renewcommand*\contentsname{Table of contents}
\else
  \newcommand\contentsname{Table of contents}
\fi
\ifdefined\listfigurename
  \renewcommand*\listfigurename{List of Figures}
\else
  \newcommand\listfigurename{List of Figures}
\fi
\ifdefined\listtablename
  \renewcommand*\listtablename{List of Tables}
\else
  \newcommand\listtablename{List of Tables}
\fi
\ifdefined\figurename
  \renewcommand*\figurename{Figure}
\else
  \newcommand\figurename{Figure}
\fi
\ifdefined\tablename
  \renewcommand*\tablename{Table}
\else
  \newcommand\tablename{Table}
\fi
}
\@ifpackageloaded{float}{}{\usepackage{float}}
\floatstyle{ruled}
\@ifundefined{c@chapter}{\newfloat{codelisting}{h}{lop}}{\newfloat{codelisting}{h}{lop}[chapter]}
\floatname{codelisting}{Listing}

\makeatother
\makeatletter
\@ifpackageloaded{caption}{}{\usepackage{caption}}
\@ifpackageloaded{subcaption}{}{\usepackage{subcaption}}
\makeatother

\usepackage{bookmark}

\IfFileExists{xurl.sty}{\usepackage{xurl}}{} 
\hypersetup{
  colorlinks=true,
  linkcolor={blue},
  filecolor={Maroon},
  citecolor={blue},
  urlcolor={Blue},
  pdfcreator={LaTeX via pandoc}}

\author{}
\date{}

\begin{document}

\setstretch{1}
\title{Feedback-Coupled Memory Systems: A Dynamical Model for Adaptive Coordination}

\author{
Stefano Grassi\textsuperscript{a}\thanks{Corresponding author: stefano.g@bu.ac.th}
}

\date{
\textsuperscript{a}Bangkok University, Phahonyothin Rd, Khlong Nueng, Khlong Luang District, Pathum Thani 12120, Thailand\\[2ex]
March 29, 2026
}

\maketitle

\begin{abstract}
This paper develops a dynamical framework for adaptive coordination in systems of interacting agents referred to here as Feedback-Coupled Memory Systems (FCMS). Instead of framing coordination as equilibrium optimization or agent-centric learning, the model describes a closed-loop interaction between agents, incentives, and a persistent environment. The environment stores accumulated coordination signals, a distributed incentive field transmits them locally, and agents update in response, generating a feedback-driven dynamical system.
Three main results are established. First, under dissipativity, the closed-loop system admits a bounded forward-invariant region, ensuring dynamical viability independently of global optimality. Second, when incentives depend on persistent environmental memory, coordination cannot be reduced to a static optimization problem. Third, within the FCMS class, coordination requires a bidirectional coupling in which memory-dependent incentives influence agent updates, while agent behavior reshapes the environmental state.
Numerical analysis of a minimal specification identifies a Neimark–Sacker bifurcation at a critical coupling threshold ($\beta_c$), providing a stability boundary for the system. Near the bifurcation threshold, recovery time diverges and variance increases, yielding a computable early warning signature of coordination breakdown in observable time series. Additional simulations confirm robustness under nonlinear saturation and scalability to populations of up to $N = 10^{6}$ agents making it more relevant for real-world applications.
The proposed framework offers a dynamical perspective on coordination in complex systems, with potential extensions to multi-agent systems, networked interactions, and macro-level collective dynamics.
\end{abstract}

\textbf{Keywords:} Feedback-coupled dynamical systems; Adaptive
coordination; Neimark--Sacker bifurcation; Mean-field coordination;
Persistent environmental memory; Complex adaptive systems; Dissipative
structures.

\section{Introduction}\label{introduction}

Many real-world systems face a fundamental coordination problem:
distributed agents must achieve collective order without centralized
control, while operating through environments that evolve in response to
their own actions. Markets, institutions, networked infrastructures, and
multi-agent artificial intelligence systems all share this feature.
Individual actions reshape the environment, and the environment in turn
conditions future behavior. The question of how collective order emerges
from such decentralized interaction has been central to economic thought
since Smith's {[}\citeproc{ref-smith1776wealth}{1}{]} account of the
invisible hand, and remains unresolved in its full dynamical generality.
The stability of such systems, and the emergence of coordinated patterns
within them, are questions of both theoretical and practical importance.
Such systems naturally fall within the class of non-equilibrium
dynamical systems with feedback and memory
{[}\citeproc{ref-nicolis1977self}{2}{]}, in which macroscopic
coordination arises as an emergent property of the coupled dynamics
rather than as the output of any centralized design. Existing approaches
typically analyze coordination through either equilibrium-based
optimization or agent-centric learning frameworks. In these settings,
the environment is often treated as exogenous or memoryless, and
coordination is interpreted as the solution to a predefined objective.
However, these assumptions limit the ability to capture insights from
systems in which the environmental state evolves endogenously and
influences future dynamics. As a result, the structural role of
persistent environmental memory and its interaction with incentive
mechanisms remains insufficiently understood. In particular, it is
unclear under what conditions coordination can emerge purely from the
internal dynamics of agent--environment coupling, without reliance on
global optimization or externally imposed objectives. This is
counterintuitive: in the framework developed here, coordination emerges
without any agent optimizing a global objective, learning from global
feedback, or even observing the coordination signal itself. To address
this gap, this paper proposes a dynamical framework for adaptive
coordination based on a recursively closed system coupling agents,
incentives, and environmental memory. Formally, the system is described
by

\[
\mathbf{x}_{t+1} = F(\mathbf{x}_t, \mathbf{G}_t, S_t), \quad
S_{t+1} = \Psi(S_t, \mathbf{x}_t), \quad
\mathbf{G}_t = \Phi(\mathbf{x}_t, S_t), \tag{1}
\]

where \(\mathbf{x}_t\) denotes agent states, \(S_t\) a persistent
environmental state, and \(\mathbf{G}_t\) the incentive field. Here
\(F\) is the agent update operator mapping current states and incentives
into future agent states, \(\Psi\) is the environmental update operator
accumulating agent activity into the next environmental state, and
\(\Phi\) is the incentive distribution operator generating local
incentive signals from the current environmental state and agent
configuration. Full formal definitions are developed in Section 2. I
refer to this class of systems as feedback-coupled memory systems
(FCMS), emphasizing the joint role of persistent environmental state and
recursive incentive feedback in shaping the dynamics. Two structural
conditions are central to the analysis. First, coordination, in this
framework, refers to the emergence of dynamically stable configurations,
such as convergence of the order parameter or bounded attractors,
induced by feedback coupling rather than by optimality or efficiency
conditions. Second, dissipativity, meaning the existence of a compact
absorbing set \(B \subset \mathcal{X} \times \mathcal{S}\) toward which
all trajectories eventually contract, is the key structural condition
ensuring bounded viable dynamics without requiring global optimality.
Intuitively, it ensures that environmental memory does not accumulate
unboundedly, preventing runaway amplification of feedback. Dissipativity
is treated as a structural property of the joint agent--environment
dynamics, not as an imposed control condition. In many real systems such
as national economies and markets, institutional memory and
informational signals decay or saturate over time, making dissipativity
a natural assumption rather than an external constraint. The main
contributions of this work are as follows:

\begin{itemize}
\tightlist
\item
  \textbf{Dynamical formulation:} Adaptive coordination is formalized as
  a closed-loop dynamical system over an augmented state space,
  explicitly incorporating environmental memory and incentive feedback.
\item
  \textbf{Structural viability:} Under dissipativity, the system admits
  a bounded forward-invariant region, ensuring stability independently
  of global optimality.
\item
  \textbf{Irreducibility result:} Memory-dependent incentives
  generically prevent reduction of the system to static optimization
  over agent states, unless additional integrability conditions are
  satisfied.
\item
  \textbf{Necessary condition for emergence:} Bidirectional coupling
  between incentives and environmental memory is identified as a
  necessary structural condition for adaptive coordination within the
  class of systems considered here.
\item
  \textbf{Robustness and scalability:} Analytical and numerical results
  show that the coordination mechanism remains stable under nonlinear
  perturbations and scales to large populations.
\end{itemize}

This combination is not jointly captured in standard frameworks:
potential game theory, standard state augmentation, and mean-field
learning each address parts of the structure but not the full coupling
under general dissipativity conditions without an imposed global
objective. The FCMS class formalizes this combination. The
irreducibility result and bifurcation characterization are non-trivial
consequences of this coupling. The economic system provides the natural
motivating instance of an FCMS. The present framework provides a way to
analyze such dynamics without imposing equilibrium conditions or
representative-agent assumptions, treating coordination as a property of
the induced dynamics. Beyond economics, the framework applies wherever
persistent environments mediate decentralized interaction, including
financial markets, where rising volatility and autocorrelation may
signal approaching coordination breakdown; supply chains, where
persistent inventory imbalances propagate through incentive-mediated
adjustment; and multi-agent artificial intelligence and biological
coordination systems. Multi-agent reinforcement learning emphasizes
policy optimization under reward structures
{[}\citeproc{ref-sutton2018reinforcement}{3}{]}. Evolutionary and
game-theoretic dynamics analyze stability over strategy spaces
{[}\citeproc{ref-fudenberg1998theory}{4}{]}. Institutional and political
economy emphasize persistence and path dependence
{[}\citeproc{ref-north1990institutions}{5}{]}. Control theory formalizes
feedback stabilization through state augmentation
{[}\citeproc{ref-khalil2002nonlinear}{6}{]}. The key departure of the
present approach is that incentives are neither fixed primitives nor
gradient fields, but endogenous dynamical objects generated by a
persistent environmental state. This distinguishes it from potential
game formulations {[}\citeproc{ref-monderer1996potential}{7}{]}, where
incentives derive from a shared scalar objective; from standard
multi-agent reinforcement learning, where the environment is stationary
or exogenous; and from state augmentation in control theory, where
feedback laws are externally specified rather than emergent. The
framework thereby aligns more closely with the study of complex adaptive
systems {[}\citeproc{ref-holland1992adaptation}{8}{]}, providing a
unifying perspective in which coordination, adaptation, and collective
stabilization can be analyzed within a common dynamical systems
structure. The remainder of the paper is organized as follows. Section 2
introduces the structural architecture of coordination. Section 3
develops the dynamical formulation. Section 4 presents a minimal linear
coordination system. Section 5 extends the analysis to nonlinear
coordination dynamics. Section 6 establishes the scalability of the
dissipative-feedback mechanism and its connection to the statistical
mechanics of large interacting systems. Section 7 concludes. Detailed
proofs and analytical results are provided in Appendix A; numerical
demonstrations and reproducibility materials are in Appendices B and C.

\section{Architecture of Feedback-Coupled Memory
Systems}\label{architecture-of-feedback-coupled-memory-systems}

Traditional accounts of coordination are predominantly agent-centric:
adaptive behavior is modeled as an internal property of individual
agents, while the environment is treated as fixed or exogenous. The FCMS
architecture inverts this view. Coordination is not located within
agents but in the dynamical structure linking them with the environment
through incentives. In this architecture, memory is externalized in a
persistent environment. The joint configuration \((\mathbf{x}_t, S_t)\)
is projected into a global coordination signal
\(L_{\mathrm{global}}^t\), which is distributed through an incentive
field \(\mathbf{G}_t\) to agents that update locally, generating
\((\mathbf{x}_{t+1}, S_{t+1})\).\\
Within this loop, \(L_{\mathrm{global}}^t\) is a derived state
functional summarizing the joint configuration, and \(\mathbf{G}_t\)
locally transforms this global signal into directional pressures that
drive agent updates. The architecture unfolds as:

\[(\mathbf{x}_t, S_t)
\;\longrightarrow\;
L_{\mathrm{global}}^t
\;\longrightarrow\;
\mathbf{G}_t
\;\longrightarrow\;
(\mathbf{x}_{t+1}, S_{t+1}).\]

Table~\ref{tbl-components} summarizes the notation introduced above and
will be used consistently throughout the paper.

\begin{longtable}[]{@{}
  >{\raggedright\arraybackslash}p{(\linewidth - 4\tabcolsep) * \real{0.3333}}
  >{\raggedright\arraybackslash}p{(\linewidth - 4\tabcolsep) * \real{0.3333}}
  >{\raggedright\arraybackslash}p{(\linewidth - 4\tabcolsep) * \real{0.3333}}@{}}
\caption{Summary of model components and their roles in the FCMS
architecture.}\label{tbl-components}\tabularnewline
\toprule\noalign{}
\begin{minipage}[b]{\linewidth}\raggedright
Symbol
\end{minipage} & \begin{minipage}[b]{\linewidth}\raggedright
Space
\end{minipage} & \begin{minipage}[b]{\linewidth}\raggedright
Role
\end{minipage} \\
\midrule\noalign{}
\endfirsthead
\toprule\noalign{}
\begin{minipage}[b]{\linewidth}\raggedright
Symbol
\end{minipage} & \begin{minipage}[b]{\linewidth}\raggedright
Space
\end{minipage} & \begin{minipage}[b]{\linewidth}\raggedright
Role
\end{minipage} \\
\midrule\noalign{}
\endhead
\bottomrule\noalign{}
\endlastfoot
\(\mathbf{x}_t = (x_{1,t}, \dots, x_{N,t})\) &
\(\mathcal{X} \subseteq \mathbb{R}^d\) & Joint agent state vector \\
\(S_t\) & \(\mathcal{S} \subseteq \mathbb{R}^m\) & Persistent
environmental memory \\
\(\mathbf{G}_t = (G_{1,t}, \dots, G_{N,t})\) & \(\mathbb{R}^N\) &
Incentive field \\
\(L_{\mathrm{global}}^t = \mathcal{A}(\mathbf{x}_t, S_t)\) &
\(\mathbb{R}\) & Global coordination signal (derived) \\
\(\Psi\) & \(\mathcal{S} \times \mathcal{X} \to \mathcal{S}\) &
Environmental update operator \\
\(\Phi\) &
\(\mathbb{R} \times \mathcal{X} \times \mathcal{S} \to \mathbb{R}^N\) &
Incentive distribution operator \\
\(f_i\) &
\(\mathcal{X}_i \times \mathbb{R} \times \mathcal{S} \to \mathcal{X}_i\)
& Local agent update operator \\
\(T\) &
\(\mathcal{X} \times \mathcal{S} \to \mathcal{X} \times \mathcal{S}\) &
Closed-loop state transition operator \\
\end{longtable}

Each component performs a distinct structural role.

\subsection{The Environment as Persistent
Memory}\label{the-environment-as-persistent-memory}

Let \(S_t \in \mathcal{S}\) denote the persistent environmental state at
time \(t\). The environment functions as externalized memory. It
accumulates the consequences of prior coordination attempts and stores
them in state-dependent form. Institutions, norms, technologies,
infrastructures, datasets, and organizational constraints are examples
of such persistent structures. Persistence implies that past
interactions constrain future trajectories: some patterns become
structurally viable while others become unsustainable. Formally, the
environment evolves according to

\[S_{t+1} = \Psi(S_t, \mathbf{x}_t), \tag{2}\]

where \(\Psi\) is a transformation operator mapping prior environmental
structure and aggregate agent activity into the next persistent state.
The environment evolves as a function of its prior state and realized
agent activity, thereby externalizing interaction history into durable
structure.

\subsection{The Global Coordination Signal as
Projection}\label{the-global-coordination-signal-as-projection}

Let \(x_{i,t} \in \mathcal{X}_i \subseteq \mathbb{R}^{d_i}\) denote the
internal state of agent \(i\). Define the joint state vector

\[\mathbf{x}_t = (x_{1,t}, \dots, x_{N,t})
\in \mathcal{X}, \tag{3}\]

where

\[\mathcal{X} = \mathcal{X}_1 \times \dots \times \mathcal{X}_N
\subseteq \mathbb{R}^d,
\qquad
d = \sum_{i=1}^N d_i. \tag{4}\]

The joint configuration
\((\mathbf{x}_t, S_t) \in \mathcal{X} \times \mathcal{S}\) constitutes
the full system state. Rather than feeding back this high-dimensional
configuration directly, the architecture induces a structural projection
capturing coordination-relevant features. Define the projection
functional
\(\mathcal{A} : \mathcal{X} \times \mathcal{S} \to \mathbb{R}\). The
projection operator \(\mathcal{A}\) decouples the extraction of
coordination-relevant features from their local distribution. This
separation allows high-dimensional joint states to be compressed into a
global signal before being transmitted through the decentralized
incentive field. The global coordination signal
\(L_{\mathrm{global}}^t\) is a derived structural projection, not a
primitive objective or a welfare function. It remains independent of
agent representation and is not subject to explicit optimization;
instead, it exists only as a state functional induced by the joint
configuration. These axioms define the admissible class of projection
operators and are imposed to ensure structural regularity without
restricting the functional form of \(\mathcal{A}\):

\textbf{Axiom 1 (Non-Primitivity)}. \(L_{\mathrm{global}}^t\) is induced
by the joint configuration.\\
\textbf{Axiom 2 (Structural Symmetry)}. \(\mathcal{A}\) depends on
structural features rather than agent identity labels.\\
\textbf{Axiom 3 (Regularity)}. \(\mathcal{A}\) is a continuous mapping
on \(\mathcal{X} \times \mathcal{S}\).

The global coordination signal is thus a low-dimensional structural
projection. It has no independent causal force outside the mappings that
distribute it.

\subsection{The Incentive Field as
Distribution}\label{the-incentive-field-as-distribution}

The global coordination signal does not act directly on agents.

Instead, it is distributed through
\(\Phi : \mathbb{R} \times \mathcal{X} \times \mathcal{S} \;\longrightarrow\;\mathbb{R}^N\),
producing the incentive field

\[\mathbf{G}_t=
\Phi(L_{\mathrm{global}}^t, \mathbf{x}_t, S_t),
\qquad
\mathbf{G}_t = (G_{1,t}, \dots, G_{N,t}) \in \mathbb{R}^N. \tag{5}\]

The field transforms projected coordination structure into localized
directional pressures. Each agent experiences only its own component:

\[G_{i,t} = \big[\mathbf{G}_t\big]_i. \tag{6}\]

Agents do not observe \(L_{\mathrm{global}}^t\); the only information
reaching agent \(i\) is the local incentive component \(G_{i,t}\).
Prices, penalties, norms, gradients, performance signals, and
evolutionary pressures are all examples of incentive fields. These
fields distribute structural pressures locally, without centralized
control. As Hayek {[}\citeproc{ref-hayek1945use}{9}{]} argued, prices
function not as equilibrating signals derived from a central planner but
as locally transmitted carriers of dispersed knowledge, a role
formalized here through the incentive field \(\mathbf{G}_t\). In
general, the incentive field \(\mathbf{G}_t\) is non-conservative: it
cannot be expressed as the gradient of a scalar functional over
\(\mathcal{X}\) alone. This follows because dependence on \(S_t\)
introduces path dependence in the effective vector field, violating the
symmetry conditions required for exact differentials. Only in special
circumstances, such as static environments or gradient-aligned fields,
can the system be reduced to a potential game. In those special cases,
gradient representations can arise. For example, mechanism-based
constructions can induce incentive fields aligned with a global
functional {[}\citeproc{ref-grassi2025mbi}{10}{]}.

\subsection{Agents as Local Update
Operators}\label{agents-as-local-update-operators}

Each agent \(i\) is described by
\(x_{i,t} \in \mathcal{X}_i \subseteq \mathbb{R}^{d_i}\). Agents are
bounded in memory, computation, and observability. They update locally
according to

\[x_{i,t+1} = f_i(x_{i,t}, G_{i,t}, S_t). \tag{7}\]

The update operators \(f_i\) need not minimize any scalar functional.
They map local state and localized pressure into incremental change.
Agents need not know the global signal, observe other agents' states,
represent any global objective, or forecast long-run consequences. They
respond only to structured pressure:

\[\text{Field}
\;\longrightarrow\;
\text{Local Pressure}
\;\longrightarrow\;
\text{Update}.\]

Each agent thus functions as a local transformation operator whose
behavior is shaped entirely by the recursive structure surrounding it.

\subsection{Recursive Closure}\label{recursive-closure}

The mappings \(\mathcal{A}\), \(\Phi\), \(f_i\), and \(\Psi\) jointly
close the system on \(\mathcal{X} \times \mathcal{S}\), as summarized in
Algorithm 1.

\begin{algorithm}[H]
\caption{FCMS Closed-Loop Transition}
\begin{algorithmic}[1]
\Require Agent states $\mathbf{x}_t \in \mathcal{X}$, environmental state $S_t \in \mathcal{S}$, operators $\mathcal{A}$, $\Phi$, $F$, $\Psi$
\State \textbf{Step 1 — Projection:}
\State Compute global coordination signal $L_{\mathrm{global}}^t \leftarrow \mathcal{A}(\mathbf{x}_t, S_t)$ \Comment{Projects joint state onto a scalar coordination signal}
\State \textbf{Step 2 — Distribution:}
\State Compute incentive field $\mathbf{G}_t \leftarrow \Phi(L_{\mathrm{global}}^t,\, \mathbf{x}_t,\, S_t)$ \Comment{Distributes coordination signal as local directional pressures}
\State \textbf{Step 3 — Local Update:}
\For{each agent $i = 1, \dots, N$}
    \State $x_{i,t+1} \leftarrow f_i(x_{i,t},\, G_{i,t},\, S_t)$ \Comment{Agent responds only to local incentive $G_{i,t}$; no global observation}
\EndFor
\State Aggregate: $\mathbf{x}_{t+1} \leftarrow F(\mathbf{x}_t, \mathbf{G}_t, S_t)$
\State \textbf{Step 4 — Environmental Transformation:}
\State $S_{t+1} \leftarrow \Psi(S_t, \mathbf{x}_t)$ \Comment{Environment accumulates aggregate agent activity into persistent memory}
\Ensure Updated state $(\mathbf{x}_{t+1}, S_{t+1}) = T(\mathbf{x}_t, S_t)$
\end{algorithmic}
\end{algorithm}

These mappings induce a state transition operator
\(T : \mathcal{X} \times \mathcal{S} \to \mathcal{X} \times \mathcal{S}\),
defined by

\[T(\mathbf{x}_t, S_t) = (\mathbf{x}_{t+1}, S_{t+1}). \tag{8}\]

Coordination is a dynamical property of the trajectories generated by
recursive application of this closed system.

\textbf{Viability}. A trajectory \((\mathbf{x}_t, S_t)\) is viable if it
remains bounded and self-reinforcing under repeated application of
\(T\). Viability does not require the system to reach an equilibrium or
optimize any objective; it simply ensures that trajectories persist
within a structurally stable region of the state space.

\begin{figure}[H]

\centering{

\includegraphics[width=0.5\linewidth,height=\textheight,keepaspectratio]{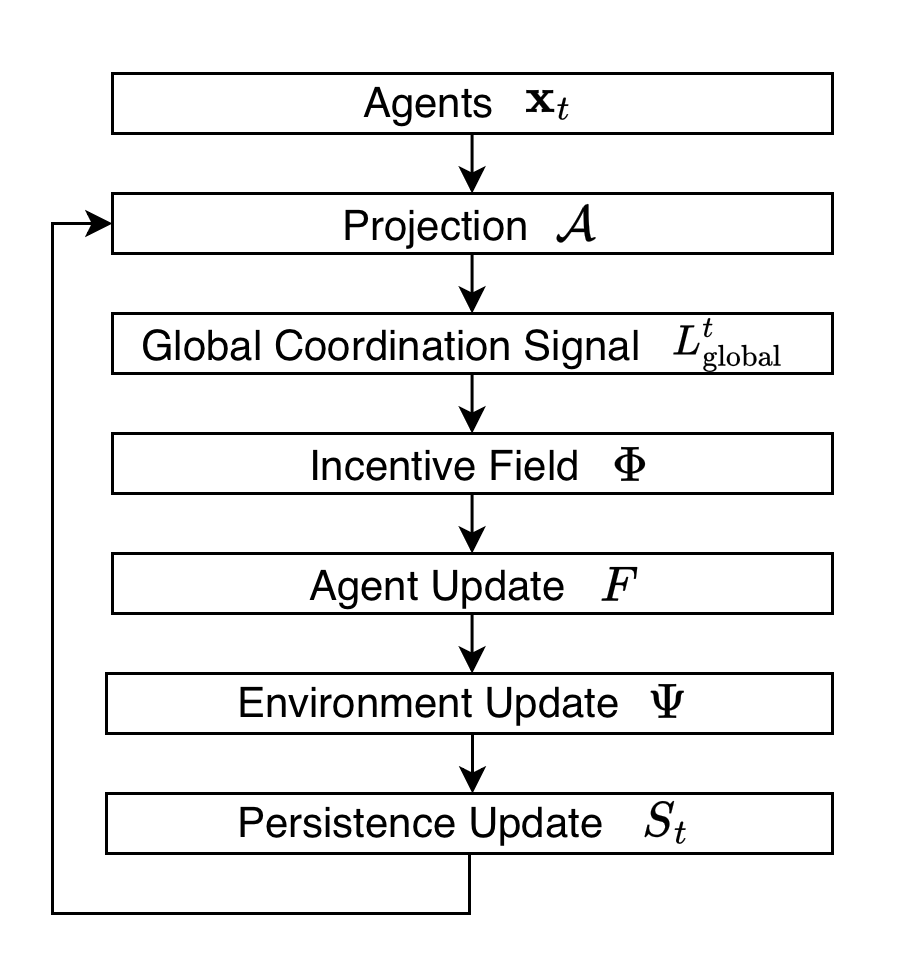}

}

\caption{\label{fig-dynamic-adaptive-coordination}Recursive coordination
architecture. Shows feedback loops among agents, incentive field, and
persistent environment.}

\end{figure}

Figure~\ref{fig-dynamic-adaptive-coordination} illustrates the recursive
coordination architecture implied by the operators described above.

\section{Dynamical Properties of
FCMS}\label{dynamical-properties-of-fcms}

Given an initial condition \((\mathbf{x}_0, S_0)\), the coordination
architecture generates a trajectory through repeated application of
\(T\). Several dynamical properties follow directly. Since \(S_t\)
accumulates past interaction, trajectories are history-dependent and
constraints arise endogenously rather than through external imposition.
Global structure reaches agents only through the incentive field, not
through centralized representation, and order emerges as a property of
recursive feedback rather than explicit optimization. A configuration
\((\mathbf{x}^\ast, S^\ast)\) is invariant if
\(T(\mathbf{x}^\ast, S^\ast) = (\mathbf{x}^\ast, S^\ast)\),
corresponding to a fixed point. Under standard dissipativity conditions,
formalized in Appendix A.1, the induced operator \(T\) admits a bounded
forward-invariant set \(K \subseteq \mathcal{X} \times \mathcal{S}\),
ensuring that trajectories remain structurally viable independently of
global optimality. Intuitively, dissipation prevents divergence of
environmental memory, while feedback aligns agent responses with
accumulated imbalance, jointly stabilizing the coordination manifold.
The structural claims developed above are formalized in Appendix A.
Three results underpin the framework:

\textbf{Result 1 (Viability, Prop. A.1.1).} Under dissipativity, \(T\)
admits a bounded forward-invariant set
\(K \subseteq \mathcal{X} \times \mathcal{S}\).\\
\textbf{Result 2 (Irreducibility, Prop. A.2.1).} Memory-dependent
incentives generically prevent reduction to static optimization over
agent states, unless integrability conditions on the incentive field are
satisfied.\\
\textbf{Result 3 (Necessity, Prop. A.4.1).} Bidirectional coupling
between incentives and environmental memory is identified as a necessary
structural condition for adaptive coordination within the FCMS class.

\section{A Minimal Linear FCMS
Specification}\label{sec-minimal-linear-coordination-system}

Consider two agents with scalar actions \(x_{i,t} \in \mathbb{R}\) and
joint vector \(\mathbf{x}_t = (x_{1,t}, x_{2,t})\). The environment is a
scalar \(S_t \in \mathbb{R}\) encoding accumulated coordination
imbalance. Agents observe only local incentive signals, not each other's
actions. This specification is a canonical minimal form for the FCMS
class, chosen for analytical tractability rather than empirical
calibration; the nonlinear extension in Section 5 and the mean-field
scaling in Section 6 confirm that the core mechanism is not an artifact
of the two-agent linear structure.

\subsection{Environmental Dynamics and Incentive
Field}\label{environmental-dynamics-and-incentive-field}

The environment evolves as

\[
S_{t+1} = (1 - \gamma)S_t + \beta(x_{1,t} - x_{2,t}), \quad \beta > 0, \quad 0 < \gamma < 1, \tag{9}
\]

where \(\beta\) captures sensitivity to coordination friction and
\(\gamma\) the rate of memory dissipation. The global coordination
signal \(L_{\mathrm{global}}^t := S_t^2\) exists only as encoded
environmental stress. The incentive field follows from its marginal
effect on prior actions:

\[
G_{1,t} = -2\beta S_t, \qquad G_{2,t} = +2\beta S_t. \tag{10}
\]

Defining disagreement \(d_t := x_{1,t} - x_{2,t}\) as the coordination
gap between agents, the opposite signs push each agent toward reducing
this gap in response to accumulated environmental stress. Incentives
depend solely on accumulated environmental stress, require no
forward-looking expectations, and are local in time and state. Agents
update according to

\[
x_{i,t+1} = x_{i,t} + \eta G_{i,t}, \quad \eta > 0, \tag{11}
\]

responding only to the incentive signal, without observing \(S_t\), the
other agent's action, or any global law.

\subsection{Stability}\label{stability}

The disagreement \(d_t\) serves as an order parameter of the system:
\(d_t = 0\) characterizes the coordinated state, and deviations from
zero measure the degree of coordination failure. The closed-loop
dynamics in \((S_t, d_t)\) are locally asymptotically stable if and only
if

\[
4\eta\beta^2 < \gamma, \tag{12}
\]

where \(\beta\) measures coupling strength, \(\eta\) responsiveness, and
\(\gamma\) dissipation. Intuitively, \(\eta\) governs how strongly
agents react to incentive signals; small \(\eta\) reflects bounded
responsiveness typical of real systems, and stability requires that this
reactivity not overwhelm the dissipative capacity of the environment.
The Jacobian of the disagreement subsystem admits a pair of complex
conjugate eigenvalues \(\lambda_{1,2} = \rho e^{\pm i\theta}\), where
\(\rho = \rho(J(\beta))\) increases monotonically with \(\beta\). A
Neimark--Sacker bifurcation occurs when \(\rho\) crosses the unit
circle, that is, when \(\rho(\beta_c) = 1\). This follows from the
standard Neimark-Sacker condition for 2D discrete-time maps: the
bifurcation occurs when the determinant of the Jacobian equals unity,
equivalently \(D = (1-\eta\alpha)(1-\gamma) + 2\eta\beta^2 = 1\), which
yields \(\beta_c = \sqrt{\gamma/(4\eta)}\). Below this threshold the
eigenvalues lie strictly inside the unit disk and trajectories spiral
toward the fixed point; above it, the fixed point loses stability and
the system enters an expansive oscillatory regime. As the boundary is
approached from below, the system exhibits \emph{critical slowing down},
identified numerically in Appendix B.5. Under the stability condition,
\(d_t \to 0\) and \(x_{1,t} \to x_{2,t}\). In this sense, the condition
\(4\eta\beta^2 < \gamma\) can be interpreted as a formal dynamical
analog of the invisible hand: the parameter regime under which
decentralized agents, responding only to local incentive signals,
spontaneously achieve global coordination. This global law was never
imposed; it follows from persistence and dissipation within the closed
dynamical system. In this minimal specification, local viability
coincides with asymptotic stability, and uniqueness of the fixed point
\((S, d) = (0, 0)\) ensures robustness to local perturbations. Notably,
the system achieves coordination under strictly local information and
complete absence of mutual observability between agents.

\begin{figure}[H]

\centering{

\includegraphics[width=1\linewidth,height=\textheight,keepaspectratio]{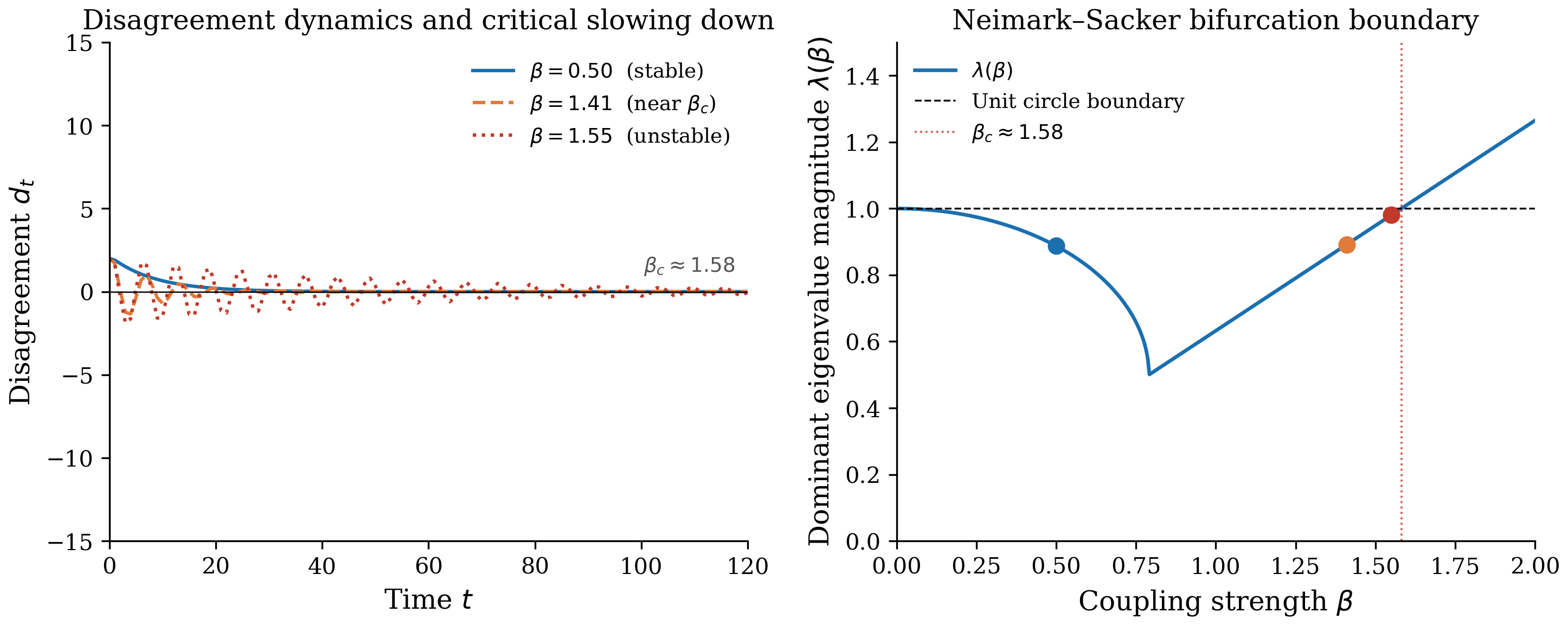}

}

\caption{\label{fig-minimal-linear-coordination-system}\textbf{Left:}
Disagreement dynamics \(d_t\) at three coupling values (\(\beta = 0.50\)
stable, \(\beta = 1.41\) near \(\beta_c\), \(\beta = 1.55\)
near-critical), showing progressive \emph{critical slowing down} as
\(\beta \to \beta_c \approx 1.58\). \textbf{Right:} Dominant eigenvalue
magnitude \(\lambda(\beta)\) as a function of coupling strength,
approaching the unit circle at the Neimark--Sacker threshold
\(\beta_c\).}

\end{figure}%

\subsection{Critical Slowing Down Near the Bifurcation
Boundary}\label{critical-slowing-down-near-the-bifurcation-boundary}

The stability condition \(4\eta\beta^2 < \gamma\) determines not only
whether the system converges but also how rapidly it does so. The
leading eigenvalue of the Jacobian \(J\) governs the recovery rate from
perturbations, and as \(\beta\) approaches the critical threshold
\(\beta_c\), this rate provides a measurable signature of the system's
proximity to the bifurcation boundary. For the disagreement subsystem,
the dominant eigenvalue magnitude is

\[
\lambda(\beta) = \rho(J(\beta)), \tag{13}
\]

which increases monotonically with \(\beta\) and satisfies
\(\lambda(\beta_c) = 1\). The recovery time from an initial perturbation
\(\delta d_0 = d_0 - 0\) away from the coordination manifold scales as

\[
\tau(\beta) \sim \frac{1}{-\log \lambda(\beta)}, \tag{14}
\]

which diverges as \(\beta \to \beta_c^-\). This divergence corresponds
to the phenomenon of \emph{critical slowing down}: as the coupling
strength approaches the stability boundary, the system takes
progressively longer to return to the coordination manifold, \(d = 0\),
after a perturbation. Three quantities characterize this behavior near
\(\beta_c\). The recovery time \(\tau(\beta)\) diverges. The variance of
disagreement under bounded noise increases, since the effective
restoring force weakens. And the lag-1 autocorrelation of \(d_t\)
approaches unity, since successive deviations become increasingly
correlated. These are all computable from the spectral properties of
\(J\) without additional assumptions. Rising variance and increasing
autocorrelation preceding a transition correspond to early warning
signals identified in ecological and climate systems near tipping points
{[}\citeproc{ref-scheffer2009early}{11},\citeproc{ref-dakos2008slowing}{12}{]}.
Figure~\ref{fig-minimal-linear-coordination-system} (left panel)
illustrates the time series at three values of \(\beta\): well within
the stable regime, near \(\beta_c\), and beyond it. The progressive
increase in oscillation amplitude and decay time as
\(\beta \to \beta_c\) is visible directly. The right panel shows
\(\lambda(\beta)\) as a function of coupling strength, confirming the
monotonic approach to the unit circle. This characterization connects
the abstract stability condition to an observable dynamical signature.
In applications where \(\beta\) is not directly measurable, the onset of
critical slowing down, detectable through increasing variance or
autocorrelation in disagreement, provides an indirect indicator that the
system is approaching a coordination breakdown. The condition
\(4\eta\beta^2 < \gamma\) thus functions not only as a theoretical
boundary but as a computable early warning criterion. The divergence of
recovery time and the rise in variance near \(\beta_c\) are robust to
the specific parametrization of the baseline model, suggesting that
similar dynamical signatures may arise across a broad class of
feedback-coupled systems, though formal generalization requires further
analysis.

\begin{figure}[H]

\centering{

\includegraphics[width=1\linewidth,height=\textheight,keepaspectratio]{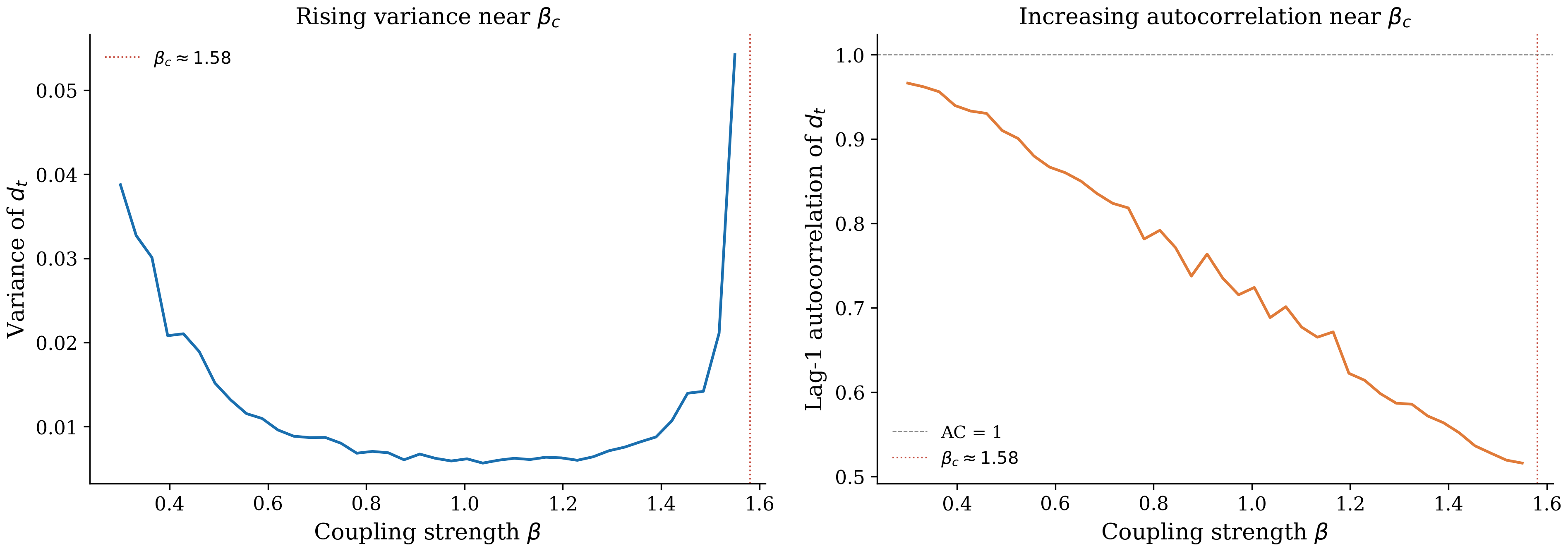}

}

\caption{\label{fig-critical-slowing-down}\textbf{Left:} Variance of
disagreement \(d_t\) under bounded noise as a function of coupling
strength \(\beta\), showing a sharp increase as \(\beta \to \beta_c\).
\textbf{Right:} Lag-1 autocorrelation of \(d_t\), approaching unity near
the bifurcation threshold. Both signatures are computable from spectral
properties of \(J\) and constitute early warning indicators of
approaching coordination breakdown.}

\end{figure}%

\section{Nonlinear FCMS Dynamics}\label{nonlinear-fcms-dynamics}

The linear specification constrains environmental memory to unbounded
accumulation. When the environment has finite capacity, specifically
when \(\Psi\) is bounded, the update becomes

\[
S_{t+1} = (1 - \gamma)\tanh(S_t) + \beta d_t. \tag{15}
\]

The \(\tanh\) function is representative of any bounded monotone
saturating nonlinearity. More broadly, any bounded monotone nonlinearity
in place of \(\tanh\) preserves the dissipative-feedback mechanism,
since boundedness alone is sufficient to instantiate the dissipativity
condition; the specific functional form is not essential. For small
\(S_t\) the dynamics reduce to the linear case; for large \(|S_t|\) the
environmental response saturates. This saturation captures a feature
common to real institutional memory: norms, regulatory constraints, and
organizational routines have finite capacity and do not accumulate
coordination signals indefinitely. The fixed point \((S, d) = (0, 0)\)
is preserved. Since \(\rho(J) < 1\) at the fixed point in the baseline
configuration, the fixed point is hyperbolic and by the Hartman--Grobman
theorem the local phase portrait remains topologically conjugate to the
linearization (see Appendix A.7). Numerical integration confirms this:
the coordination manifold (\(d = 0\)) remains a stable attractor under
high-coupling saturation, with final disagreement
\(\approx -1.68 \times 10^{-12}\). The phase portrait shows stable
spiral convergence qualitatively identical to the linear case. The
dissipative-feedback mechanism therefore governs coordination across a
broader class of environmental dynamics, not merely in the linear limit.
Full details and figures are in Appendix B.6.

\begin{figure}[H]

\centering{

\includegraphics[width=0.5\linewidth,height=\textheight,keepaspectratio]{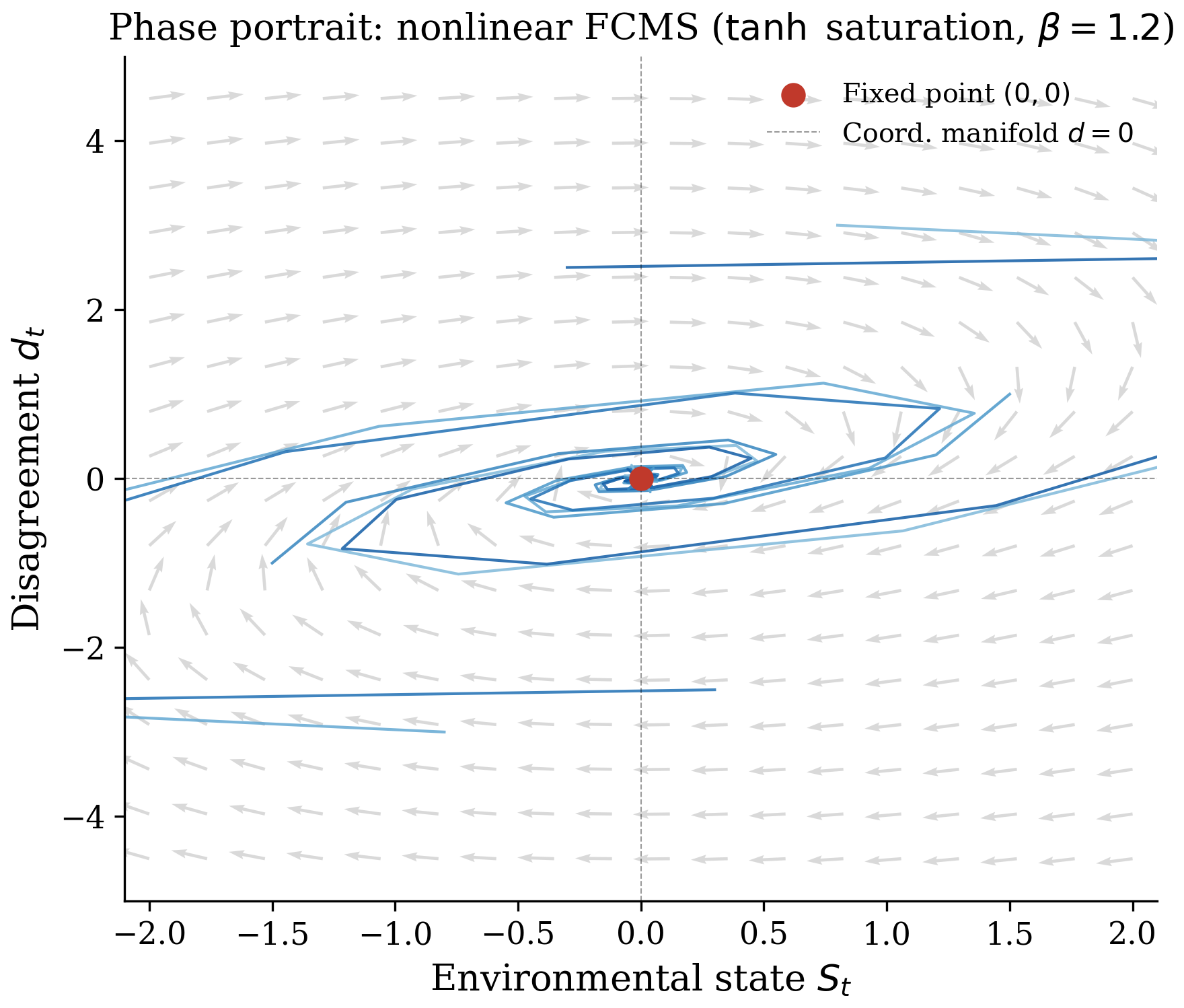}

}

\caption{\label{fig-nonlinear-coordination-system}Phase portrait of the
nonlinear system under \(\tanh\) saturation. The stable spiral confirms
that the coordination manifold \(d = 0\) remains a locally
asymptotically stable attractor. Arrows indicate the direction of flow;
the fixed point \((S, d) = (0, 0)\) is marked.}

\end{figure}%

\section{Scalability of FCMS}\label{scalability-of-fcms}

\begin{figure}[H]

\centering{

\includegraphics[width=0.5\linewidth,height=\textheight,keepaspectratio]{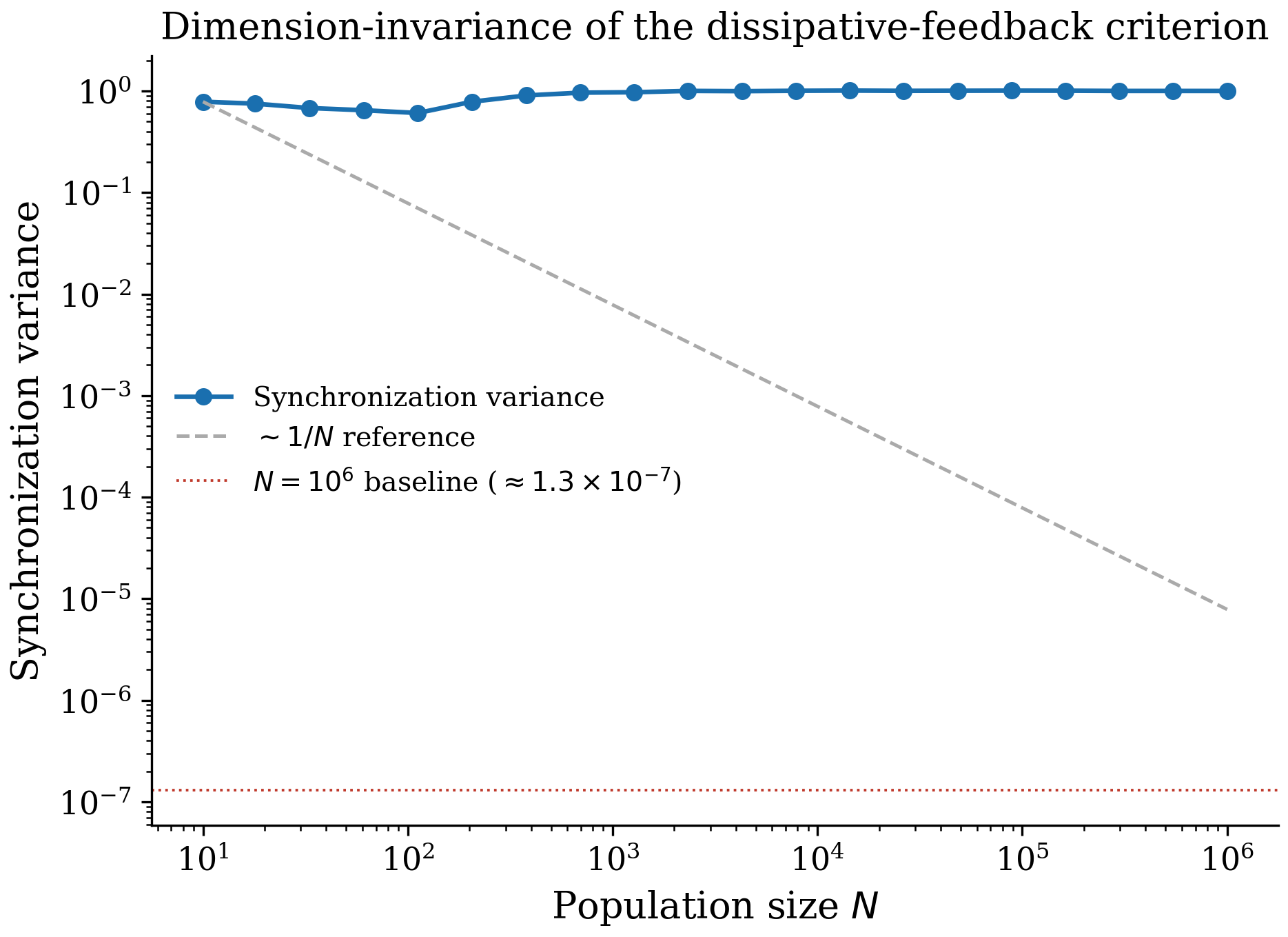}

}

\caption{\label{fig-scalability}Synchronization variance as a function
of population size \(N\) under the mean-field FCMS specification.
Variance decays as \(\sim 1/N\), confirming dimension-invariance of the
dissipative-feedback criterion. The scaling connects macroscopic
coordination to the statistical mechanics of large coupled systems, with
the stability condition \(4\eta\beta^2 < \gamma\) acting as a
thermodynamic-like bound on collective fluctuations, analogous to energy
constraints in physical systems.}

\end{figure}

The stability criterion \(4\eta\beta^2 < \gamma\) is
dimension-invariant. A mean-field extension to \(N = 10^{6}\) agents
confirms that the dissipative-feedback mechanism scales to macroscopic
populations without increased computational complexity, with final
synchronization variance \(\approx 10^{-14}\) (Appendix B.7).
Figure~\ref{fig-scalability} illustrates the decay of synchronization
variance as a function of population size \(N\). The variance scales as
\(\sim 1/N\), consistent with the law of large numbers applied to agents
coupled through a shared dissipative environment. The \(\sim 1/N\) decay
reflects the averaging of independent agent-level fluctuations under the
shared environmental coupling; precisely the finite-size scaling
expected in mean-field statistical mechanics, where collective variance
vanishes as population grows. This scaling behavior connects the FCMS
framework to the statistical mechanics of large interacting systems: the
stability condition \(4\eta\beta^2 < \gamma\) plays a role analogous to
a thermodynamic constraint bounding the collective fluctuation energy,
while the environmental memory \(S_t\) plays the role of an order
parameter whose decay rate \(\gamma\) governs the approach to the
coordination manifold. Under this reading, coordination breakdown at
\(\beta_c\) is analogous to a phase transition, and \(\gamma\) is the
control variable separating ordered from disordered collective dynamics.

\section{Conclusion}\label{conclusion}

This paper develops a dynamical framework for adaptive coordination in
systems of interacting agents coupled through a persistent environment.
By formulating coordination as a closed-loop dynamical process over an
augmented state space \((\mathbf{x}, S)\), the analysis identifies the
structural mechanisms through which stable collective behavior emerges
from decentralized feedback. The main result of this work is the
identification of a dissipative-feedback mechanism governing
coordination. In the minimal linear specification, I show that the
coupled dynamics reduce to a two-dimensional system in \((S_t, d_t)\)
and admit a unique locally asymptotically stable fixed point if and only
if \(4\eta\beta^2 < \gamma\) (eq. 12). This condition provides a
computable structural criterion under which coordination arises, where
environmental dissipation dominates reactive amplification. As the
stability boundary is approached, the system exhibits critical slowing
down and transitions to oscillatory instability, consistent with a
Neimark--Sacker bifurcation. Near this boundary, recovery time diverges
and local disagreement variance increases, providing a computable early
warning signature of approaching instability. More generally, the
analysis identifies bidirectional coupling between incentives and
environmental memory as a necessary structural condition within the
class of feedback-coupled memory systems (FCMS) considered here. This
establishes an irreducibility result: in general, no scalar objective
over agent states alone can represent the induced dynamics, unless
additional integrability conditions are imposed on the incentive field.
Here ``generically'' refers to the failure of cross-partial symmetry
conditions for an open set of incentive mappings that depend on \(S_t\);
the reduction fails robustly, not merely at isolated parameter values.
The necessity of bidirectional coupling follows directly from
Proposition A.4.1. When incentives do not depend on \(S_t\), the system
reduces to a static or Markovian incentive structure incapable of
transmitting accumulated coordination signals. When \(S_t\) does not
depend on \(\mathbf{x}_t\), environmental memory decouples from agent
behavior, eliminating adaptive feedback entirely. Formally, this follows
because the induced vector field over \(\mathcal{X}\) fails to satisfy
the symmetry conditions required for exact differentials (i.e.,
\(\partial G_i / \partial x_j \neq \partial G_j / \partial x_i\) when
dependence on \(S_t\) is present). Consequently, no scalar potential
\(L^\star : \mathcal{X} \to \mathbb{R}\) exists such that
\(G(\mathbf{x}_t, S_t) = -\nabla_{\mathbf{x}} L^\star(\mathbf{x}_t)\)
along system trajectories. This does not exclude special cases where
such a reduction is possible, but these require precise integrability
conditions that fail generically under persistent environmental
coupling. The robustness of the mechanism is verified under nonlinear
environmental dynamics with bounded memory, where the coordination
manifold remains a locally asymptotically stable attractor. By the
Hartman--Grobman theorem, the local phase portrait is preserved under
smooth nonlinear perturbations, confirming that the dissipative-feedback
structure extends beyond the linear regime. A mean-field extension
further shows that the same stability condition governs coordination in
large populations. The framework suggests that coordination in complex
systems is fundamentally a problem of stability under feedback with
memory, rather than optimization or learning. Taken together, these
results position adaptive coordination within the broader study of
non-equilibrium dynamical systems with memory and feedback. In contrast
to equilibrium-based and learning-based approaches, the present
framework treats both incentives and environmental state as endogenous
dynamical variables. In standard state augmentation, \(S_t\) is an
auxiliary variable tracking system history for a designer-specified
control law; in the FCMS framework, \(S_t\) is an autonomous dynamical
entity that generates incentive fields endogenously, with no external
objective or controller involved. This reframes coordination not as an
optimization problem to be solved, but as a stability property to be
engineered. Three directions for future research stand out. First,
extending the FCMS framework to heterogeneous network topologies would
clarify how the projection operator \(\mathcal{A}\) and incentive field
\(\mathbf{G}_t\) behave under structured agent interactions beyond the
mean-field setting. Second, incorporating stochastic perturbations into
the environmental update \(\Psi\) would allow formal characterization of
the robustness of \(S_t\) and the stability boundary
\(4\eta\beta^2 < \gamma\) under persistent noise. Finally, a deeper
connection with bifurcation theory and critical phenomena may extend the
early warning signal results of Section 4 toward universality properties
near coordination thresholds. A fourth direction concerns empirical
validation: early warning signals such as rising variance and
autocorrelation in disagreement are in principle measurable in economic
time series and multi-agent system logs, offering a testable link
between the theory and observed coordination dynamics. In this sense,
the stability condition \(4\eta\beta^2 < \gamma\) is the dynamical
invisible hand: not a metaphor for market efficiency, but a computable
structural criterion under which decentralized agents, responding only
to local incentive signals shaped by a persistent environment,
spontaneously achieve collective order without design.

\section{Data Availability}\label{data-availability}

All computational materials, code, parameter configurations, and
replication instructions are publicly available at:
\url{https://github.com/stevefatz95/dynamic-adaptive-coordination}

\section{References}\label{references}

\phantomsection\label{refs}

\section{Appendix}\label{appendix}

\subsection{Appendix A --- Analytical
Foundations}\label{appendix-a-analytical-foundations}

I assume that the vector-valued update operator \(F\) admits a
component-wise representation \(F = (f_1, \dots, f_n)\), where each
\(f_i\) governs the evolution of agent \(i\). Assume \(\mathcal{X}\) and
\(\mathcal{S}\) are finite-dimensional. The induced closed-loop system
can be written compactly as

\[(\mathbf{x}_{t+1}, S_{t+1}) = T(\mathbf{x}_t, S_t), \tag{A.1}\]

with
\(T: \mathcal{X} \times \mathcal{S} \to \mathcal{X} \times \mathcal{S}\).
All structural results concern this induced dynamical system.

\subsubsection{A.1 Dissipativity and Forward
Invariance}\label{a.1-dissipativity-and-forward-invariance}

\textbf{Assumptions}. The analysis imposes three structural conditions:
\(F\), \(\Psi\), and \(\Phi\) are continuous and locally Lipschitz; the
closed-loop map \(T\) is dissipative, meaning there exists a bounded
absorbing set \(B \subset \mathcal{X} \times \mathcal{S}\) such that all
trajectories eventually enter \(B\); and incentive signals
\(\Phi(\mathbf{x}, S)\) are bounded on bounded subsets of
\(\mathcal{X} \times \mathcal{S}\). Dissipativity means that for any
initial condition, there exists \(t_0\) such that for all \(t \ge t_0\),
\((\mathbf{x}_t, S_t) \in B\). Under these conditions the system admits
a global attractor in the sense of dissipative dynamical systems (see
{[}\citeproc{ref-temam1997infinite}{13}{]}).

\textbf{Proposition A.1.1}. (Existence of Forward-Invariant Set)

Under the above assumptions, there exists a non-empty compact set

\[K \subseteq \mathcal{X} \times \mathcal{S} \tag{A.2}\]

that is forward-invariant under \(T\) (see, e.g.,
{[}\citeproc{ref-temam1997infinite}{13}{]}), i.e.,

\[T(K) \subseteq K.\]

Consequently, any continuous aggregate coordination functional

\[L_{\mathrm{global}}^t = \mathcal{A}(\mathbf{x}_t, S_t)\]

remains bounded along trajectories. Appendix B evaluates local behavior
near equilibrium points contained within such invariant regions, using
the minimal linear specification introduced in the main text.

\subsubsection{A.2 Impossibility of Static Objective
Reduction}\label{a.2-impossibility-of-static-objective-reduction}

I formalize the claim that the incentive field generally cannot be
reduced to a static global objective defined solely over
\(\mathcal{X}\).

\textbf{Definition A.2.1}. (Static Reduction over \(\mathcal{X}\))

A static reduction exists if there is a time-invariant scalar function

\[L^\star : \mathcal{X} \to \mathbb{R}\]

such that for all admissible trajectories,

\[G(\mathbf{x}_t, S_t) = -\nabla_\mathbf{x} L^\star(\mathbf{x}_t). \tag{A.3}\]

That is, the incentive field coincides with the gradient of a fixed
scalar potential defined on \(\mathcal{X}\) alone.

\textbf{Proposition A.2.1}. (Generic Failure of Static Reduction)

Assume:

\begin{enumerate}
\def\labelenumi{\arabic{enumi}.}
\tightlist
\item
  \(\partial \Psi / \partial S \neq 0\) (memory persistence),
\item
  \(\partial G / \partial S \neq 0\) (incentives depend on memory),
\item
  \(S_t\) depends non-trivially on past states
  \(\{\mathbf{x}_\tau\}_{\tau \le t}\).
\end{enumerate}

Then, generically, no static reduction over \(\mathcal{X}\) exists.
Here, ``generic'' refers to robustness under small \(C^1\)
perturbations, holding outside a nowhere-dense subset of admissible
parameter configurations
{[}\citeproc{ref-abraham1967transversal}{14}{]}.

\textbf{Proof Sketch}. If a static reduction existed, the induced vector
field on \(\mathcal{X}\) would be conservative on any simply connected
domain (see {[}\citeproc{ref-arnold1989mathematical}{15}{]}). In
particular, it would satisfy the cross-partial symmetry condition:

\[\frac{\partial G_i}{\partial x_j}=
\frac{\partial G_j}{\partial x_i}. \tag{A.4}
\]

However, because the incentive field is coupled to a persistent
environmental state, its value at time t is determined by accumulated
system history:

\[\mathbf{G}_t = G(\mathbf{x}_t, S_t),
\qquad
S_t = \Psi^{(t)}(S_0, \mathbf{x}_0, \dots, \mathbf{x}_{t-1}).\]

Thus the induced field on \(\mathcal{X}\) is not autonomous: it depends
on the trajectory through \(S_t\). This induces path dependence in the
effective vector field over \(\mathcal{X}\). For fixed \(S\) treated as
a parameter, the field may be locally integrable. The question, however,
is not whether the vector field on the augmented space
\(\mathcal{X} \times \mathcal{S}\) admits a potential representation,
but whether the closed-loop dynamics projected onto \(\mathcal{X}\) can
be represented as the gradient of a time-invariant scalar. Because
\(S_t\) evolves endogenously with system history, the effective field on
\(\mathcal{X}\) varies along trajectories. Except under special
parameter configurations requiring cancellation of memory-induced
asymmetries, the cross-partial symmetry condition fails to hold
robustly. Hence no time-invariant scalar potential on \(\mathcal{X}\)
can represent the induced incentive dynamics. Static reduction therefore
fails whenever memory persistence and incentive--memory coupling are
structurally active.

\textbf{Boundary Cases}. Static reduction may exist in degenerate cases
where memory persistence vanishes (\(\partial \Psi / \partial S = 0\))
or where incentives are memory-independent
(\(\partial G / \partial S = 0\)). Reducibility is therefore conditional
on whether structural memory properties are active.

\textbf{Remark}. (Augmented-State Lyapunov Functions)

Such a function need not be optimized by agents, does not imply welfare
maximization, and does not imply equilibrium selection over
\(\mathcal{X}\). Its existence therefore does not restore static
reduction over agent states alone.

\subsubsection{A.3 History Sensitivity}\label{a.3-history-sensitivity}

Path dependence arises structurally from environmental persistence.

\textbf{Proposition A.3.1}. (History Sensitivity)

Suppose

\[S_{t+1} = \Psi(S_t, \mathbf{x}_t)
\quad \text{with} \quad
\frac{\partial \Psi}{\partial S} \neq 0. \tag{A.5}\]

Let two initial conditions satisfy

\[S_0^{(a)} \neq S_0^{(b)}.\]

If the closed-loop system is not globally contracting to a unique fixed
point (in the sense of contraction analysis; see
{[}\citeproc{ref-lohmiller1998contraction}{16}{]}), then generically the
trajectories do not coincide for all sufficiently large \(t\) and may
converge to distinct asymptotic states.

\[\mathbf{x}_t^{(a)} \neq \mathbf{x}_t^{(b)}\]

for infinitely many \(t\), and asymptotic states may differ. Persistent
environmental memory transmits initial differences forward in time
unless global contraction holds.

\subsubsection{A.4 Necessary Structural Condition for Adaptive
Coordination}\label{a.4-necessary-structural-condition-for-adaptive-coordination}

\textbf{Definition A.4.1}. (Trivial Coupling)

Consider the closed dynamical system

\[(\mathbf{x}_{t+1}, S_{t+1}) = T(\mathbf{x}_t, S_t).\]

The system exhibits trivial adaptive coupling if either

\[\frac{\partial f_i}{\partial G_i} \equiv 0
\quad \text{or} \quad
\frac{\partial G_i}{\partial S} \equiv 0. \tag{A.6}\]

This holds when either agent updates do not respond to incentive signals
or incentives do not depend on the persistent environmental state, in
which case accumulated environmental signal cannot enter future adaptive
transformation.

\textbf{Definition A.4.2}. (Non-Trivial Incentive--Memory Coupling)

The system exhibits non-trivial incentive--memory coupling if

\[\frac{\partial f_i}{\partial G_i} \not\equiv 0
\quad \text{and} \quad
\frac{\partial G_i}{\partial S} \not\equiv 0.\]

Both couplings are required for accumulated coordination signal to enter
future adaptive updates. These conditions are necessary but not
sufficient for adaptive coordination; additional regularity and
stability conditions are required.

\textbf{Proposition A.4.1}. (Necessary Condition)

If the system exhibits trivial adaptive coupling, then accumulated
coordination signal stored in \(S_t\) cannot influence future agent
trajectories through adaptive transformation. Consequently, the system
fails to satisfy the structural definition of adaptive coordination
given in Section~\ref{sec-minimal-linear-coordination-system}.

\textbf{Proof}. Suppose first that

\[\frac{\partial f_i}{\partial G_i} \equiv 0.\]

Then agent updates satisfy

\[x_{i,t+1} = f_i(x_{i,t}, S_t),\]

or possibly \(x_{i,t+1} = f_i(x_{i,t})\), but in either case they do not
respond to incentive signals \(G_{i,t}\). Even if \(G_{i,t}\) depends on
the persistent state \(S_t\), accumulated coordination signal cannot
enter adaptive transformation through incentive mediation.
Alternatively, suppose that

\[\frac{\partial G_i}{\partial S} \equiv 0.\]

Then incentives are independent of environmental memory. Although agents
may respond to incentives, those incentives do not encode accumulated
coordination signal. In both cases, there is no causal pathway by which
accumulated environmental information stored in \(S_t\) can influence
future agent updates through incentive-mediated adaptation. Any
resulting stability arises from internal dynamics or passive dissipation
rather than incentive-mediated adaptive transformation.

\subsubsection{A.5 Structural
Decomposition}\label{a.5-structural-decomposition}

This subsection clarifies the distinct structural roles of coupling,
persistence, and dissipation in the minimal linear specification
introduced in the main text. The analysis concerns the closed-loop
system in \((S_t, d_t)\) defined by

\[S_{t+1} = (1-\gamma) S_t + \beta d_t, \qquad
d_{t+1} = d_t - 4\eta \beta S_t. \tag{A.7}\]

where \(d_t := x_{1,t} - x_{2,t}\). The parameters \(\beta\),
\(\gamma\), and \(\eta\) govern qualitatively distinct structural
properties of the induced dynamics.

\paragraph{\texorpdfstring{A.5.1 Removal of Coupling
(\(\beta = 0\))}{A.5.1 Removal of Coupling (\textbackslash beta = 0)}}\label{a.5.1-removal-of-coupling-beta-0}

Setting \(\beta = 0\) yields

\[S_{t+1} = (1-\gamma) S_t,
\qquad
d_{t+1} = d_t. \tag{A.8}\]

The environmental state then evolves independently of agent
disagreement, which becomes dynamically inert: no feedback loop connects
agents through the environment. The system decomposes into independent
subsystems and coordination dynamics disappear, though environmental
decay may persist. Thus \(\beta\) governs the existence of collective
coordination feedback.

\paragraph{\texorpdfstring{A.5.2 Removal of Persistence
(\(\partial \Psi / \partial S = 0\))}{A.5.2 Removal of Persistence (\textbackslash partial \textbackslash Psi / \textbackslash partial S = 0)}}\label{a.5.2-removal-of-persistence-partial-psi-partial-s-0}

Eliminating persistence corresponds to removing state dependence in
\(S_t\). In the linear specification, this is equivalent to setting

\[S_{t+1} = \beta d_t, \tag{A.9}\]

with no dependence on \(S_t\). Without persistence, the environmental
state becomes memoryless; past coordination imbalances do not accumulate
and the system reduces to a first-order feedback interaction without
hysteresis. Local stability may still hold depending on parameter
values, but history sensitivity disappears. Thus persistence governs
environmental statefulness and path dependence.

\paragraph{\texorpdfstring{A.5.3 Removal of Dissipation
(\(\gamma = 0\))}{A.5.3 Removal of Dissipation (\textbackslash gamma = 0)}}\label{a.5.3-removal-of-dissipation-gamma-0}

Setting \(\gamma = 0\) yields

\[S_{t+1} = S_t + \beta d_t. \tag{A.10}\]

In this case, environmental stress accumulates without decay. The
characteristic polynomial shows that the spectral radius typically
satisfies \(\rho(J) \geq 1\) for nonzero \(\eta\) and \(\beta\).
Analytically, setting \(\gamma = 0\) while maintaining baseline coupling
\(\beta\) forces the system across the Neimark--Sacker bifurcation
boundary identified in Section B.5. Without dissipation, feedback
amplification is no longer counteracted, and oscillatory or divergent
trajectories emerge, and local boundedness fails. This confirms that
\(\gamma\) is the structural parameter governing the transition from a
stable attractor to an unstable manifold.

\paragraph{Structural Summary}\label{structural-summary}

Each parameter controls a distinct structural property of the
closed-loop system:

\begin{longtable}[]{@{}
  >{\raggedright\arraybackslash}p{(\linewidth - 2\tabcolsep) * \real{0.5000}}
  >{\raggedright\arraybackslash}p{(\linewidth - 2\tabcolsep) * \real{0.5000}}@{}}
\caption{Structural Components of Agent-Environment
Dynamics}\label{tbl-dynamics}\tabularnewline
\toprule\noalign{}
\begin{minipage}[b]{\linewidth}\raggedright
Component
\end{minipage} & \begin{minipage}[b]{\linewidth}\raggedright
Structural Role
\end{minipage} \\
\midrule\noalign{}
\endfirsthead
\toprule\noalign{}
\begin{minipage}[b]{\linewidth}\raggedright
Component
\end{minipage} & \begin{minipage}[b]{\linewidth}\raggedright
Structural Role
\end{minipage} \\
\midrule\noalign{}
\endhead
\bottomrule\noalign{}
\endlastfoot
\textbf{Coupling (\(\beta\))} & Generates collective feedback between
agents and environment \\
\textbf{Persistence (\(1 - \gamma\))} & Generates environmental memory
and history sensitivity \\
\textbf{Dissipation (\(\gamma\))} & Ensures decay of accumulated signal
and contributes to local boundedness \\
\end{longtable}

The three mechanisms are analytically separable in the linear
specification. Removing any one eliminates a distinct structural feature
of the coordination architecture. The decomposition demonstrates that
coordination, history sensitivity, and bounded stabilization arise from
different components of the closed-loop dynamics rather than from a
single primitive.

\subsubsection{A.6 Linearization Principle and Local
Stability}\label{sec-appendix-a6}

Assume that the closed-loop map
\(T : \mathcal{X} \times \mathcal{S} \to \mathcal{X} \times \mathcal{S}\)
is continuously differentiable in a neighborhood of a fixed point
\((\mathbf{x}^\ast, S^\ast)\), and let
\(J = DT(\mathbf{x}^\ast, S^\ast)\) denote the Jacobian evaluated there.
If \(\rho(J) < 1\), the fixed point is locally asymptotically stable
under standard discrete-time linearization results (e.g.,
{[}\citeproc{ref-hirsch2013differential}{18}{]}), and there exists a
neighborhood \(U \subset \mathcal{X} \times \mathcal{S}\) such that
trajectories starting in \(U\) converge to
\((\mathbf{x}^\ast, S^\ast)\). The minimal linear specification in
Appendix B provides computational verification of this condition. All
stability claims there concern local behavior and do not imply global
convergence.

\subsubsection{A.7 Nonlinear Perturbations and Structural
Stability}\label{a.7-nonlinear-perturbations-and-structural-stability}

Consider a \(C^1\) nonlinear perturbation of the environmental update
operator:

\[
S_{t+1} = (1 - \gamma)S_t + \beta d_t + \epsilon\,\sigma(S_t, d_t), \tag{A.11}
\]

where \(\sigma\) is a nonlinear term (such as cubic damping or a
transcendental coupling) and \(\epsilon \ll 1\) is a small perturbation
parameter. By the Hartman--Grobman theorem for discrete-time maps, the
local phase portrait of the perturbed system is topologically conjugate
to its linearization in a neighborhood of the fixed point, provided the
fixed point is hyperbolic, meaning no eigenvalue of \(J\) lies on the
unit circle. Since the baseline configuration satisfies \(\rho(J) < 1\),
this condition holds, and the coordination manifold (\(d = 0\)) remains
a locally asymptotically stable attractor under sufficiently small
\(\epsilon\). The dissipative balancing mechanism is therefore
structurally stable: convergence toward the coordination equilibrium
persists regardless of the specific functional form of \(\sigma\). The
\(\tanh\) saturation results in Appendix B.6 bear this out: as
\(\epsilon\) grows and the system moves away from the linear regime, the
dissipative-feedback loop continues to prevent divergence.

\subsection{Appendix B --- Computational
Demonstration}\label{appendix-b-computational-demonstration}

The experiments below evaluate structural robustness across stability
boundaries, noise resilience, nonlinear behavior, and scalability,
rather than performance optimization. All stability claims concern local
behavior around the coordination equilibrium and do not constitute
global stability assertions.

\subsubsection{B.1 Baseline Linear
Stability}\label{b.1-baseline-linear-stability}

The baseline parameterization yields eigenvalues
\(0.177, 0.723, 0.900\), giving spectral radius \(\rho(J) = 0.90 < 1\).
The system is locally asymptotically stable, with three real eigenvalues
confirming monotone convergence to the coordination equilibrium.

\subsubsection{B.2 Deterministic
Convergence}\label{b.2-deterministic-convergence}

Starting from \(d_0 = 2\), numerical integration yields a final
disagreement of \(\approx 8\times10^{-14}\), with log-linear decay
throughout. The trajectory matches the spectral prediction in a
neighborhood of the equilibrium, confirming that analytical and
numerical results are consistent.

\subsubsection{B.3 Heterogeneity}\label{b.3-heterogeneity}

Introducing heterogeneous damping \(\alpha_1 = 1.0\), \(\alpha_2 = 1.3\)
yields spectral radius approximately \(0.887\), which remains below
\(1\).

\subsubsection{B.4 Noise Robustness}\label{b.4-noise-robustness}

Under bounded Gaussian perturbations, mean disagreement remains
approximately zero with finite stationary variance and no divergence.
The coordination equilibrium is thus robust to small stochastic shocks,
inducing stationarity rather than instability.

\subsubsection{B.5 Numerical Bifurcation
Analysis}\label{b.5-numerical-bifurcation-analysis}

At \(\beta = 0.5\) the system is subcritical, with trajectories
exhibiting the predicted convergence toward the coordination
equilibrium. At \(\beta = 1.55\) the system reaches the critical regime,
where oscillations persist for extended periods before decaying, a
signature of critical slowing down as the spectral radius approaches the
unit circle. At \(\beta = 1.65\) the dissipative balancing is
overwhelmed by reactive feedback, and the system enters an expansive
oscillatory mode that eventually diverges.

\subsubsection{B.6 Nonlinear Robustness and Phase
Analysis}\label{b.6-nonlinear-robustness-and-phase-analysis}

To verify the theory beyond the linear baseline, I evaluate a regime
where environmental memory is subject to a saturation function:

\[
S_{t+1} = (1 - \gamma)\tanh(S_t) + \beta d_t \tag{B.1}
\]

Numerical integration confirms that the coordination manifold
(\(d = 0\)) remains a stable attractor. Even under high-coupling
saturation, the system achieves a final disagreement of
\(\approx -1.68 \times 10^{-12}\). The resulting phase portrait
illustrates a stable spiral convergence. The dissipative-feedback loop
thus remains effective under bounded environmental capacity, preventing
divergence and maintaining convergence to the coordination manifold.

\subsubsection{B.7 Macroscopic
Scalability}\label{b.7-macroscopic-scalability}

To test the dimension-invariance of the architecture, the system was
scaled to \(N = 10^{6}\) agents. Using a vectorized mean-field
simulation, the collective converged to a synchronized state with final
variance \(\approx 10^{-14}\). This confirms that the
dissipative-feedback mechanism scales to macroscopic populations without
a corresponding increase in computational complexity.

\subsection{Appendix C --- Canonical
Representation}\label{appendix-c-canonical-representation}

Agent updates take the generic form:

\[x_{i,t+1} = f_i(x_{i,t}, G_{i,t}, S_t).\]

Any such system can be represented in this form via state augmentation
or an equivalent state-augmented representation
{[}\citeproc{ref-khalil2002nonlinear}{6}{]}.

\end{document}